\documentstyle[12pt]{article}

\begin{document}

\title {GUT's have the ability to defy sphalerons}
\author{SUBIR MOHAN \\
       The Mehta Research Institute \\
           10, Kasturba Gandhi Marg\\
           (Old Kutchery Road)\\
            ALLAHABAD - 211002 (U.P.) INDIA }
\maketitle

\centerline{\large \bf Abstract}


We show that the baryon asymmetry produced by the out of equilibrium decay
of heavy $GUT$ scalars can be the baryon asymmetry that is observed today.
No restrictions need be imposed on the initial values of $B$, $L$ and
$B-L$, nor on the neutrino masses; no new symmetries need be gauged nor
new fermions introduced. We find two mechanisms that can bring this about
for any $GUT$ gauge group. Two illustrative models are discussed that are
robust, and need just two (at most four) more $GUT$ scalar fields than the
minimal model. The additional scalar fields can also help in generating an
adequately large value of the CP violation parameter and in efficiently
annihilating the monopoles. Our work should firmly re-establish heavy
$GUT$ scalars as probable progenitors of today's baryon asymmetry.

\newpage

The baryon (lepton) asymmetry $B(L)$, with $B-L=0$, generated well above
the electroweak phase transition temperature $(T_{EW})$ is known to be
erased by the electroweak instanton (sphaleron) interactions $[1]$. A
non-zero value of $B-L$ invariably comes hand in hand with massive
neutrinos and to prevent the lepton number violating interactions, such
as decay of heavy right handed Majorana neutrinos, from being in
thermodynamic equilibrium at the same time as the sphaleron interactions,
a combination that can potentially wipe out baryon asymmetry with any
value of $B-L$, special restrictions have to be imposed on the neutrino
masses $[2]$. This state of affairs was largely responsible for a waning of
interest, over the years, in the grand unified theory $(G U T)$
based baryogenesis schemes $[3]$ and fostered the emergence of low
temperature baryogenesis scenarios, most notably electroweak baryogenesis $[4]$.

Electroweak baryogenesis schemes, basically, attempt to harness the baryon
number violating sphaleron interactions to produce a baryon asymmetry
during the course of electroweak phase transition (EWPT). If the phase
transition is sufficiently strongly first order then the expanding bubble
walls provide the site for the departure from thermal equilibrium. $C$ and
$CP$ violation are largely model dependent. Attractive though the idea is,
a large body of concerted effort has not been able to establish its
viability beyond reasonable doubt. An abundance of models that claim to
produce a baryon asymmetry in the range allowed by nucleosynthesis
constraints exists, and yet the debate on whether the mechanism is
fundamentally correct and well understood rages on $[5]$. While waiting
for the dust to clear on the issue of electroweak baryogenesis, some
workers have begun devising strategies to protect the baryon asymmetry
generated at high temperatures $(>> T_{EW})$ from sphaleron depredations.

Lew and Riotto $[6]$ introduce mirror fermions to render baryon and lepton
number currents anomaly free and then gauge the baryon and lepton number
symmetries. When $U(1)_B$ and $U(1)_L$ are broken, cosmic string loops
arise that ultimately collapse producing fermions. Since the baryon number
is anomaly free, the sphalerons leave it unaffected. Campbell et al. $[7]$
have an interesting idea. They observe that if there is no mixing of
lepton generations and lepton number violating interactions are in thermal
equilibrium for just one or two (but not all three) generations
simultaneously with the sphaleron interactions, then a non-zero value of
$B$ can survive even if initially $B-L=0$; but ${1/3} B -L_i$ or ${2/3} B -
(L_i+L_j)$ must be non-zero where i(and j) is (are) the generation(s) for
which the lepton number violating interactions are not in thermal
equilibrium. Dreiner and Ross $[8]$ have shown that, for a second or
weakly first order EWPT, inclusion of the particle masses while analyzing
the chemical equilibrium equations for $T< T_{EW}$ gives $B \sim 10^{-7} \Delta
L$ for $(B-L)_{\rm initial} =0$, where $ \Delta L$ is the initial lepton
generations' asymmetry $\Delta L = \sum\limits_{i>j}(L_i -L_j)$ and it is expected
to be of the same order of magnitude as the initial lepton number. While
Davidson et al. $[9]$ show that the inclusion of thermal mass effects for
$T{^> _{\! \sim}} T_{EW}$ also gives  $B \sim 10^{-7} \Delta L$ at $ T \sim T_{EW}$.
Thus, irrespective of the order (first, second or weakly first) of EWPT a
non-zero, though severly diluted, value of $B$ should survive today 
if initially, at $T > > T_{EW}$, $ B-L$ had been ${\rm zero.}^{\# 1}$

\vspace{.4cm}

{\bf 1.} We find that a non-zero $B$ can survive the sphaleron interactions
if the electromagnetic charge $Q$ carried by the particles in chemical
equilibrium is not zero. To maintain the electrical charge neutrality of
the universe, a charge $-Q$ can be carried by the particles
that are not in chemical ${\rm equilibrium.}^{\# 2}$ The {\it \mbox{ value
of $(B-L)$ is immaterial}}, only

$$ (B-L) \neq \frac {N Q}{(4N+ 2m)}$$

as

$$ B= \frac{(8 N+4 m)(B-L) - 2 NQ} {(22 N + 13 m)},$$

$$L= \frac{-(14 N + 9 m) (B-L) - 2 NQ} {(22 N +13 m)} \ . \eqno(1) $$

\noindent $N(m)$ is the number of standard model type fermion generations (Higgs doublets).

If lepton number violating interactions are in thermal equlibrium for all
the generations, $(B-L)$ is not conserved but 

$$B= \frac{2 NQ} {(10 N + 3m)}$$

$$ L= \frac {-6 NQ}{(10 N + 3m)} \eqno(2) $$

\noindent And if only $n$ (but not all $N$) generations have lepton number
violations in thermal equilibrium then, ${(B-L)}_{N-n} \equiv
{\displaystyle \frac{(N-n)B}{N}}
 - \sum\limits_{i}^{(N-n)} L_i$ is conserved and

$$ B = \frac {4 N[(4N + 2m){(B-L)}_{N-n} - (N-4n) Q]} {(4N + 2m)(13 N -4n)
- 8N (N - 4n)} $$
$$L  = \frac {-9 B} {4} - \frac {N(Q- 2B)}{(4N + 2m)} \eqno(3)$$

These ${\rm observations}^{\# 3}$ follow from a straightforward exercise
in solving the chemical equilbrium equations, {\it\mbox{ a la}} Harvey and Turner
$[13]$. The particles we have considered are $N$ standard model type
fermion generations, $m$ standard model type Higgs doublets and the
standard model gauge bosons. Right handed neutrinos may also be present,
but in the presence of Majorana mass terms their chemical potential is
zero. The particles that stay out of equilibrium will
usually be of the heavy $GUT $ scalar type.

The main point of this paper is to show that the baryon asymmetry produced
by the decay of heavy $GUT $ scalars(X) can be the baryon asymmetry
observed today. And the models that make this possible are simply a
modified version of the model $[14]$ that gives rise to a large value of
the CP violation parameter $(\epsilon)$ or the model $[15]$
for efficient annihilation of magnetic monopoles: a feature that makes our
main point particularly alluring. No extra symmetries need be gauged, no
new fermions added, no special restrictions imposed on neutrino masses and
the baryon asymmetry at $T > T_{EW}$ need not be large.

The mechanisms that bring about this happy turn of events are two:

(i) heavy $GUT$ scalars(X) may decay during a phase of temporarily broken
electromagnetic gauge invariance, ${U(1)}_{em}$, producing not just a
non-zero $B$ but also a non-zero $Q$, and

(ii) an asymmetry may be produced in the numbers of charged heavy $GUT$
scalars(X) which may decay at different times producing, again, a non-zero
$B$ and a non-zero $Q$.

We, first, cursorily deal with an implementation of mechanism (i) and then
give details of a model that executes mechanism (ii).

\vspace{.4cm}

{\bf 2.a.} In $[15]$ we had presented a model for efficient annihilation
of magnetic monopoles, which is achieved by temporarily breaking
${U(1)}_{em}$ for  \linebreak $ 10^7 GeV { ^<_{\! \sim}} T {^ <_{\! \sim}} 10^8 GeV$. Below $10^7 GeV$ the
monopoles begin dominating the energy density of the universe, and in a
matter dominated universe the monopoles almost cease to annihilate $[16]$.
The monopole annihilation is expected to yield $GUT$ gauge bosons and
$GUT$ scalars which can rapidly decay into fermions.

In our model, $U(1)_{em}$ is broken by the non-zero thermal expectation
value of a charged but $SU(3)_c \times SU(2)_L$ - singlet scalar which may
couple to leptons $[17]$. This gives rise to lepton number and charge
violating mass terms during the broken $U(1)_{em}$ epoch, thereby
facilitating emergence of a non-zero $Q$ and $B-L$ when the heavy $GUT$
scalar bosons, produced by monopole annihilations, rapidly decay out
of equilibrium.

Upon restoration of $U(1)_{em}$ at $T {^ <_{\! \sim}} 10^7 GeV$ a charge $-Q$ emerges
out of the vacuum, if the universe is finite sized $[18]$, in the form of
charged scalar particles (mass $\sim 10^6 GeV$) responsible for breaking
$U(1)_{em}$ and the universe regains charge neutrality. These scalar
particles may eventually decay into leptons at $T<T_{EW}$, making $B-L=0$
but leaving $B$ unaffected.

In $[19]$ the significance of a non-zero $Q$ had not yet dawned on us,
and we were quite satisfied with being able to obtain $B-L \neq 0$ for
$T{^> _{\! \sim}} T_{EW}$ in cases, such as $ SU(5)$, where it would otherwise be
zero. ${}^{\# 4}$

It should now be clear that the $GUT$ gauge group can be other than
$SU(5)$ and our model will still help preserve $B$ generated in the course
of monopole annihilations and subsequent decay of heavy $GUT$ scalar
bosons, even if lepton number violating interactions are present.

If the reheat temperature $(T_{RH})$ after inflation is less than
$T_{GUT}$, then the monopoles are simply inflated away. But we can still
arrange a temporarily broken $U(1)_{em}$ phase around $T_{RH}$ enabling
the $GUT$ scalar bosons produced by inflaton decay to decay into fermions
with $Q \neq 0$ thus allowing a non-zero $B$ to persist at $ T<T_{EW}$.

{\bf 2.b.} Finally, a model that is known to give rise to a large value of
the $CP$ violation parameter $(\epsilon)$ in the course of decay of the
heavy $GUT$ scalars is shown to be capable of protecting the baryon
asymmetry, generated by the scalar decays, from sphaleron interactions.
For the ease of presentation we shall stick to $SU(5)$, but we stress that
analogous models can be constructed for any other $GUT$ gauge group.

The Weinberg (Three - Higgs) model $[14]$ consists of, eponymous, three
scalar fields $\phi_i$ in the fundamental ${\bf 5} $ representation of
$SU(5)$, with interactions
$$ V(\phi) = \mu_{r}^{2} (\phi_{r}^{\dagger} \phi_r) + a_{rs} (\phi_{r}^{\dagger}
\phi_{r}) (\phi_{s}^{\dagger} \phi_{s})+ b_{rs}( \phi_{r}^{\dagger} \phi_{s})
(\phi_{s}^{\dagger} \phi_{r}) + c_{rs} (\phi_{r}^{\dagger} \phi_{s}) (\phi_{r}^{\dagger}
\phi_{s}) \ , \eqno(4)$$
where $r, s$ are summed from $1$ to $3$. Hermiticity requires that
$a_{rs}$ and $b_{rs}$ be real and symmetric, and $c_{rs}$ be Hermitian.
$\phi_3$ is chosen not to couple to fermions, but $\phi_1$ and $\phi_2$
have Yukawa couplings such as 

$$ L_{Yuk} = {\bar \psi}_{m, \alpha \beta} {\chi}_{n}^{\alpha}
[({f}_{1}^{\dagger})_{mn} {\phi}_1^{\beta} + ({g}_{2}^{\dagger})_{mn} {\phi}_2^{\beta}] +
\epsilon_{\alpha \beta \mu \nu \lambda} {\psi}_{m}^{\alpha \beta} C
{\psi}_{n}^{\mu \nu} [{(g_1)}_{mn} {\phi}_1^{\lambda} + {(f_2)}_{mn}
{\phi}_2^{\lambda}] + h.c.  \eqno(5) $$
where $ m, n$ are summed over fermion generations, $C$ is the
charge-conjugation matrix; and fermions are put in a right-handed
five-dimensional representation $\chi^\alpha$ and a left-handed,
ten-dimensional representation  $\psi^{\alpha \beta}$; $ \alpha, \  \beta,
\  \mu, \  \nu, \  \lambda$ are $SU(5)$ indices running from $1$ to $5$.

When $SU(5)$ is broken by the vacuum expectation value $(\sim  10^{15} GeV)$
of a scalar field in the adjoint ${\bf 24}$ representation, the color
triplets $\phi_{i}^{a}$ acquire superheavy masses which we choose to
satisfy $M_3 > M_1 > M_2$; a is the color index. The ${SU(2)}_L$ doublets
${\phi}_{3}^{h}$ and ${\phi}_{2}^{h}$ are also allowed to acquire heavy
masses $(m_3 {^> _{\! \sim}} m_2)< M_i$. ${\phi}_{1}^{h}$ becomes the standard Higgs
doublet whose vacuum expectation value $( \sim 10^2 GeV)$ gives mass to
the fermions.

The superheavy color triplet ${\phi}_{3}^{a}$ has two decay channels:
${\phi}_{1}^{a}$ plus $\bar{\phi}_{3}^{h}$ and ${\phi}_{1}^{h}$ or
${\phi}_{2}^{a}$ plus $\bar{\phi}_{3}^{h}$ and ${\phi}_{2}^{h}$.
The decays of ${\phi}_{3}^{a}$ will violate CP invariance because of the
complex $c_{rs}$. The interference between tree and one-loop diagrams
gives, for $b_{rs} << c_{rs}$,

$$ r_{3 \rightarrow 1} - {\bar r}_{3 \rightarrow 1} \equiv \epsilon = \frac
{{\rm Im} (c_{12} c_{23} c_{31})} {(4 \pi) {|c_{13}|}^2 + {|c_{23}|}^2}   \eqno(6)$$
where $r{({\bar r})}_{3 \rightarrow 1 } $ is the partial decay rate of
$\phi_{3}^{a} \bar{(\phi_{3}^{a})} $ into $\phi_{1}^{a} \bar{
(\phi_{1}^{a})} $ plus relatively lighter Higgs. In fact $\epsilon$ is the
number of $\phi_{1}^{a}  $ produced in the decay of a $\phi_{3}^{a}- 
\bar{\phi_{3}^{a}} $ pair. CPT ensures that  $r_{3 \rightarrow 2} - {\bar r}_{3
\rightarrow 2}$ is $ - \epsilon$, hence the number of $\bar{
(\phi_{2}^{a})} $'s produced in the decay of a $\phi_{3}^{a}- 
\bar{(\phi_{3}^{a})} $ pair is also $\epsilon$.

Since some of the Yukawa couplings ${(g_1)}_{mn}$ may be as large as
$O(1)$, ${\phi}_{1}^{a}$ readily decays into fermions producing

$$ B-L = 3(-2/3) \ , \eqno(7) $$
where we have summed the contributions of the three $\phi_{1}^{a}$'s. The
largest of the Yukawa couplings, ${(f_2)}_{mn}$ and ${(g_2)}_{mn}$, may be
such that $\bar{\phi_{2}^{a}} $ does not decay until well after
$T_{EW}$ when we expect the sphaleron interactions to have dropped out of
thermal equilibrium.

If initially $T{^> _{\! \sim}} T_{GUT}$, $\phi_{3}^{a} \bar{(\phi_{3}^{a})} $
begin decaying when the temperature $T_3$ has fallen below $M_3$ if, ${({\rm for
}\  c_{31} \sim c_{32})}^{\# 5}$.

$$M_3{^> _{\! \sim}} (2\times 10^{14}) {|c_{31}|}^2 GeV \ . \eqno(8)$$
Then $\phi_{1}^{a}$ decays into fermions at (say) $T_1 {^ <_{\! \sim}} M_1<M_3$, and
if $T_3 < M_1$ then $\phi_{1}^{a}$ and $\bar{{\phi}_{2}^{a}}$ produced in
$\phi_{3}^{a}$ decay are never in 
{\it{\mbox chemical equilibrium.}}
$\bar{{\phi}_{2}^{h}}, {({\phi}_{3}^{h}, \ \bar{{\phi}_{3}^{h}})}$ will
be in chemical equilibrium with fermions if $m_2(m_3) < T_1$ due to scalar mixing
through $c_{12}$. Now the charge carried by
the particles in chemical equilibrium ${\rm is}^{\# 6}$ 

$$ Q= -1 \eqno(9) $$
and it is balanced by the charge carried by $\bar{{\phi}_{2}^{a}}$,
which is $3(+1/3)$. At $T \sim m_3 > m_2$, $\phi_{3}^{h}$ and $\bar{{\phi}_{3}^{h}}$
annihilate. For $m_2 <T <m_3$ the particles in chemical equilibrium are
$N(=3)$ fermion generations, $m(=2)$ Higgs doublets ${\phi_{1}^{h}}$ and
$\bar{{\phi}_{2}^{h}}$, and the standard model gauge bosons. At $T \sim m_2$

$$B=-29/46 \hspace{1cm} L=63/46$$
$$ \mu_0 = 3/184 \hspace{1cm} Q=-1 \eqno(10)$$  
where $\mu_0$ is the chemical potential of the Higgs doublets.

At $T {^ <_{\! \sim}} m_2$, ${\phi}_{2}^{h}$ drops out of chemical equilibrium and we
are left with just one Higgs doublet in chemical equilibrium. For $T_{EW} {^ <_{\! \sim}} T < m_2$

$$B= -4609/7268, \hspace{.5cm} L= 9927/7268, \hspace{.5cm} Q= -181/184 \ .\eqno(11)$$
Sometime after the sphalerons have dropped out of thermal equilibrium, $\bar{{\phi}_{2}^{a}}$
and $\bar{{\phi}_{2}^{h}}$ decay into fermions giving 

$$B_2=1-r/2, \hspace{.5cm} L_2= -1-r/2, \hspace{.5cm} Q_2= 181/184 \eqno(12)$$
where $r \equiv 4g_{2}^{2}/(4g_{2}^{2} + 3f_{2}^{2}); m,n$ have been
summed over fermion generations. Now the net values are

$$B=\frac{2659}{7268} - \frac{r}{2}, \hspace{.5cm} L= \frac{2659}{7268}
-\frac{r}{2}, \hspace{.5cm}  Q=0 \eqno(13)$$

If ${\phi}_{3}^{a} {( \bar{{\phi}_{3}^{a}})}$ had been produced at $T<m_2$
by the collapse or annihilation of a topological defect or decay of the
inflaton then  $\bar{{\phi}_{2}^{h}}$ is never in chemical equilibrium and
${\phi}_{1}^{a}$ readily decays with 
 
$$ B-L=-2, \hspace{.5cm} Q=0 \hspace{.2cm},$$
$$B=-56/79, \hspace{.5cm} L=102/79 \ .\eqno(14)$$
And when  $\bar{{\phi}_{2}^{a}}$ and  $\bar{{\phi}_{2}^{h}}$ decay at
$T<T_{EW}$, the net values are 

$$ B=\frac{23}{79} - \frac{r}{2}, \hspace{.5cm} L=\frac{23}{79}
-\frac{r}{2}, \hspace{.5cm} Q=0 \ . \eqno(15)$$
${(f_2, g_2)}_{mn}$, $r$ can always be such that $B\neq 0$.

A few words on particle masses and decays.

(i) $M_1 > 10^{11} GeV$ to prevent too rapid a proton decay $[20]$. For
${\phi}_{1}^{a}$ to decay out of equilibrium $M_1 {^> _{\! \sim}} 5 \times 10^{15} GeV$.
But in the Weinberg model out of equilibrium decay of ${\phi}_{1}^{a}$ is
possible for smaller values of $M_1$ if $M_1>T_3$.

(ii) The values of $B(L)$ in all the equations are per  ${\phi}_{1}^{a}{(
\bar{{\phi}_{2}^{a}})}$ particle decay; and since $B \sim 10^{-1}$ the
number density of  ${\phi}_{1}^{a}{({\phi}_{2}^{a})}$ should be $
{\displaystyle \left(\frac{n_3}{s}\right)} \epsilon \sim 10^{-10}$ to allow $(n_B/s)$ to lie in the
range $(4-6) \times 10^{-11}$ allowed by nucleosynthesis. $n_3$ is the
number density of  ${\phi}_{3}^{a}{(\bar{{\phi}_{3}^{a}})}$ that decay.

(iii) To prevent too large an entropy production when
$\bar{{\phi}_{2}^{a}}$ and $\bar{{\phi}_{2}^{h}}$ decay at $1 MeV {^<_{\! \sim}}T<T_{EW}$,
$M_2 {^ <_{\! \sim}} 10^{12} GeV$. (Recall $m_2<M_2)$. $M_2$ can be larger but then
${\displaystyle \left(\frac{n_3}{s}\right)} \epsilon$ would have to be correspondingly large to accommodate
the increase in entropy. And, roughly, $(m_2, M_2) >
10^2 GeV$ or else $\phi_2$ should have been, more or less, seen by the
present accelerators.

(iv) For $M_2(m_2)$ in the range allowed by (iii), $ {(f_2, g_2)}_{mn}$
can easily be chosen to avoid too large a rate of proton decay and flavour
changing neutral current processes. Roughly, for $(M_2, m_2) > 10^6 GeV$
and ${(f_2, g_2)}_{mn} <10^{-10}$ there are no problems.

Just by weakly coupling the scalar field $\phi_2$ to fermions, the
Weinberg (Three-Higgs) model, for generating a large value of CP violation
parameter $\epsilon$, has been empowered with the ability to protect
baryon asymmetry, generated well above $T_{EW}$, from sphaleron
interactions. It should be noted that even if $B=0$ at $T {^> _{\! \sim}} T_{EW}$
(say $\bar{\phi}_{2}^{h}$ is never in chemical equilibrium but rapid
lepton number violating interactions are present while sphalerons are in
thermal equilibrium) an adequately large baryon asymmetry can still be
produced by the out of equilibrium decays of $\bar{\phi}_{2}^{a}$ at
$T<T_{EW}$ (and if such is the case then today $B-L=2$). A very robust
model for baryogenesis we have indeed.

Yukawa couplings as small as $10^{-10}$ should not seem very
unusual, for in the Standard Model the electron Yukawa coupling is
$2\times 10^{-6}$ and should neutrinos turn out to have Dirac masses in
the eV range Yukawa couplings of order $10^{-10} - 10^{-11}$ will be needed.

{\bf 3.} A simple solution to a vexing problem has been found. It appears
that today's observed baryon asymmetry is as likely to have been produced
by the decay of heavy $GUT$ scalars as by any other viable mechanism.

The traditional $GUT$'s appear to have an inherent ability to fight off
the sphalerons. All that the minimal models need are two (at the most
four) additional $GUT$ scalar fields to be adequately empowered : a very
economical bargain. And the additional scalar fields do not just shield
the baryon asymmetry from sphalerons but also eliminate monopoles or
generate and adequately large value of the CP violation parameter : a two
for one kind of offer that makes an already economical bargain even more attractive.

After a long, and now seen to be undeserved, exile to the margins,
baryogenesis via decay of heavy $GUT$ scalars seems set to regain its
position on the main stage.

\newpage

\centerline{\large \bf FOOTNOTES}

\begin{description}

\item{[\bf 1]} Cline et al. $[10]$ had proposed that an asymmetry in the
number of right-handed electrons $e_R$ could protect the baryon asymmetry.
But detailed calculations $[11]$ belied this hope by revealing that $e_R$
enter chemical equilibrium, at $T{^ <_{\! \sim}} 10 TeV$, well before the sphalerons
have dropped out of thermal equilibrium. Antaramian et al. $[12]$ also
discuss survival of baryon asymmetry but without any models.

\item{[\bf 2]} A particle species is said to be in chemical equilibrium if the
rate $(\Gamma)$ of the reactions that alter its number is large enough to
keep it in thermal equilibrium, $\Gamma >> H$.

\item{[\bf 3]} Eqs. $(1) -(3)$ pertain to $T{^> _{\! \sim}} T_{EW}$. If EWPT is second or
weakly first order, sphalerons may remain in thermal equilibrium upto $T
\sim m_W<T_{EW}$ and the numerical coefficients in eqs.$(1) -(3)$ will be
different, but qualitatively our results remain unaffected.

\item{[\bf 4]} The ability of non-zero $Q$ to protect $B$ from being decimated
by the sphalerons becomes fully manifest only in the presence of rapid
lepton-number violating interactions for all the generations.

\item{[\bf 5]} The ${\phi}_{3}^{a} - \bar {{\phi}_{3}^{a}}$ annihilations can be
ignored (at $T_3 {^ <_{\! \sim}} M_3$) if $\Gamma_{ann} < H$ which holds for $T> 3
\times 10^{14} GeV$.

\item{[\bf 6]} Total (charge, baryon number, lepton number) is
${\displaystyle \left(\frac{n_3}{s}\right)}
\epsilon (Q, B, L)$, where $n_3$ is the number density of ${\phi}_{3}^{a}
(\bar{{\phi}_{3}^{a}}) $ that decay.

\end{description}

\newpage

\centerline{\large \bf REFERENCES}

\begin{description}

\item{[\bf 1]} G. 'tHooft, Phys. Rev. Lett. {\bf 37}(1976)8; Phys. Rev. {\bf
D14}(1976)3432; \\ N.S. Manton, {\it ibid.} {\bf 28}(1983)2019; F.R.
Klinkhamer and N.S. Manton, {\it ibid.} {\bf 30}(1984)2212; \\ V. Kuzmin, V.
Rubakov and M. Shaposhnikov, Phys. Lett. {\bf B155}(1985)36; \\ P. Arnold and
L. McLerran, Phys. Rev. {\bf D37}(1988)1020.

\item{[\bf 2]} M. Fukugita and T. Yanagida, Phys. Rev. {\bf
D42}(1990)1285; \\
J.A. Harvey and M.S. Turner, {\it ibid.} 3344; \\ M. Luty, {\it ibid.} {\bf 45}(1992)455.

\item{[\bf 3]} E.W. Kolb and M.S. Turner, Annu. Rev. Nucl. Part Sci. {\bf 33}(1983)645.

\item{[\bf 4]} A.G. Cohen, D.B. Kaplan and A.E. Nelson, Annu. Rev. Nucl.
Part. Sci. {\bf 43}(1993)27.

\item{[\bf 5]} N. Turok, in Perspectives in Higgs Physics, ed. G. Kane
(World Scientific, 1992).

\item{[\bf 6]} H. Lew and A. Riotto, Phys. Rev. {\bf D49}(1994)3837.

\item{[\bf 7]} B.A. Campbell, S. Davidson, J. Ellis and K.A. Olive, Phys.
Lett. {\bf B297}(1992)118.

\item{[\bf 8]} H. Dreiner and G.G. Ross, Nucl. Phys. {\bf B410}(1993)188.

\item{[\bf 9]} S. Davidson, K. Kainulainen and K.A. Olive, Phys. Lett. {\bf
B335}(1994)339.

\item{[\bf 10} J.M. Cline, K. Kainulainen and K.A. Olive, Phys. Rev. Lett.
{\bf 71}(1993)2372.

\item{[\bf 11]} J.M. Cline, K. Kainulainen and K.A. Olive, Phys. Rev. {\bf D49}(1994)6394.

\item{[\bf 12]} A. Antaramian, L.J. Hall and A. Rasin, Phys. Rev. {\bf D49}(1994)3881.

\item{[\bf 13]} J.A. Harvey and M.S. Turner, Phys. Rev. {\bf D42}(1990)3344.

\item{[\bf 14]} S. Weinberg, Phys. Rev. Lett. {\bf 37}(1976)657; \\
             S. Barr, G. Segre and H.A. Weldon, Phys. Rev. {\bf D20}(1979)2494.

\item{[\bf 15]} S. Mohan, Mod. Phys. Lett. {\bf A10}(1995)227.

\item{[\bf 16]} E. Gates, L.M. Krauss and J. Terning, Phys. Lett. {\bf B284}(1992)309.

\item{[\bf 17]} A. Zee, Phys. Lett. {\bf B93}(1980)389; {\it ibid.} {\bf
161}(1985)141; \\ S. Mohan, as in $[19]$.

\item{[\bf 18]} T.D. Lee, Particle Physics and Introduction to Field
Theory (Harwood Academic, New York, 1981).

\item{[\bf 19]} S. Mohan, Baryogenesis in $SU(5)$ can Survive Sphaleron
Effects, MRI Preprint (October 1995), submitted for publication.

\item{[\bf 20]} E.W. Kolb and M.S. Turner, The Early Universe
(Addison-Wesley, Redwood City, CA, 1990).

\end{description}

\end{document}